\documentclass[twocolumn,prl,aps,epsfig,superscriptaddress,showpacs,twocolumn]{revtex4-1}
\usepackage{bm}
\usepackage{amsmath}
\usepackage{amssymb}
\usepackage{graphicx}
\usepackage{esint}
\usepackage[unicode=true,pdfusetitle,
 bookmarks=false,
 breaklinks=false,pdfborder={0 0 1},colorlinks=false]{hyperref}
\usepackage{bm}
\usepackage{amsfonts}
\usepackage{mathrsfs}

\hypersetup{
 colorlinks,citecolor=red}
\makeatletter

\@ifundefined{textcolor}{}
{
 \definecolor{BLACK}{gray}{0}
 \definecolor{WHITE}{gray}{1}
 \definecolor{RED}{rgb}{1,0,0}
 \definecolor{GREEN}{rgb}{0,1,0}
 \definecolor{BLUE}{rgb}{0,0,1}
 \definecolor{CYAN}{cmyk}{1,0,0,0}
 \definecolor{MAGENTA}{cmyk}{0,1,0,0}
 \definecolor{YELLOW}{cmyk}{0,0,1,0}
}
\makeatother

\begin{document}

\title{Systematic Construction of tight-binding Hamiltonians for Topological
Insulators and Superconductors}
\date{\today}
\author{D.-L. Deng$^{1,2}$, S.-T. Wang$^{1,2}$, and L.-M. Duan}
\affiliation{Department of Physics, University of Michigan, Ann Arbor,
Michigan 48109, USA}
\affiliation{Center for Quantum Information, IIIS, Tsinghua University, Beijing 100084,
PR China}

\begin{abstract}
A remarkable discovery in recent years is that there exist various kinds of
topological insulators and superconductors characterized by a periodic table
according to the system symmetry and dimensionality. To physically realize
these peculiar phases and study their properties, a critical step is to
construct experimentally relevant Hamiltonians which support these
topological phases. We propose a general and systematic method based on the
quaternion algebra to construct the tight binding Hamiltonians for all the
three-dimensional topological phases in the periodic table characterized by
arbitrary integer topological invariants, which include the spin-singlet and
the spin-triplet topological superconductors, the Hopf and the chiral
topological insulators as particular examples. For each class, we calculate
the corresponding topological invariants through both geometric analysis and
numerical simulation.
\end{abstract}

\pacs{73.43.-f, 74.20.-z, 03.65.Vf, 02.40.Re}
\maketitle

Topological insulators (TIs) and superconductors (TSCs) are symmetry
protected topological phases of non-interacting fermions described by
quadratic Hamiltonians \cite{hasan2010colloquium, *qi2011topological}, which
have robust gapless boundary modes protected by the system symmetry \cite%
{chen2013symmetry}. These protected boundary modes have exotic properties
and in some cases are characterized as anyons with fractional statistics
\cite{nayak2008non}, which could be used for the realization of topological
quantum computation \cite{kitaev2003fault}. Notable examples of TIs include
the integer quantum Hall states \cite{sarma2008perspectives} and the
recently discovered two-dimensional (2D) quantum spin Hall states \cite%
{bernevig2006quantum, *konig2007quantum} and the three-dimensional (3D) $%
\mathbb{Z}_{2}$ TIs \cite%
{fu2007topological3DTI,*roy2009z,*moore2007topological,*hsieh2008topological}%
. Examples of TSCs include the 2D $p+ip$ superconductors of spinless
fermions \cite{read2000paired} and the Helium superfluid B-phase \cite%
{kitaev2009periodic,ryu2010topological}.

It turns out that the above TI/TSC examples are just a part of a larger
scheme: they sit inside a periodic table for TIs/TSCs developed according to
symmetry and dimensionality of the system \cite%
{ryu2010topological,schnyder2008classification,kitaev2009periodic}. The
periodic table predicts possible existence of a number of new topological
phases, and it is of great interest to search for these new phases in
nature. However, the periodic table does not tell where to look for or how
to realize these phases. To physically realize these exotic phases and study
their properties, it is of critical importance to construct tight-binding
Hamiltonians so that they could be realized in real quantum materials such
as optical lattice systems \cite{bloch2012quantum}. So far, some clever
example Hamiltonians have been found for a few new topological phases \cite%
{qi2009time,roy2008topological,schnyder2008classification,moore2008topological}%
, typically with the topological invariant $\Gamma =\pm 1$, but we lack a
systematic method to construct tight-binding Hamiltonians for generic
topological phases with arbitrary integer topological invariants.

In this paper, we propose a general and systematic method to construct tight
binding Hamiltonians for new topological phases based on the use of
quaternion algebra. By this method, we construct the tight-binding
Hamiltonians for all the 3D topological phases in the periodic table with
arbitrary integer topological invariants, which include the spin-singlet and
the spin-triplet TSCs, the chiral and the Hopf TIs as prototypical examples.
For each class, the topological invariant is explicitly calculated for the
constructed Hamiltonian, using both geometric analysis and numerical
simulation, which confirm that we indeed realize all the topologically
distinct phases in the corresponding class characterized by a topological
invariant of arbitrary integer values. The construction method proposed here
should also work for the 2D and 1D cases, and we believe its direct
generalization to Clifford algebra should provide a powerful tool to
construct tight-binding Hamiltonians for all the integer topological phases
in the periodic table.

Before showing the method, let us first briefly introduce the quaternion
algebra $\mathbb{H}$, which is a generalization of the familiar complex
algebra, with the imaginary basis-vectors extended from one ($\boldsymbol{i}$%
) to three ($\boldsymbol{i},\boldsymbol{j},\boldsymbol{k}$). The
basis-vectors ($\boldsymbol{i},\boldsymbol{j},\boldsymbol{k}$) multiply
according to the following non-commutative product table \cite%
{nakahara2003geometry}:
\begin{eqnarray}
&\boldsymbol{i}^{2}=\boldsymbol{j}^{2}=\boldsymbol{k}^{2}=-1,\qquad &%
\boldsymbol{ij}=-\boldsymbol{ji}=\boldsymbol{k},  \notag \\
&\boldsymbol{jk}=-\boldsymbol{kj}=\boldsymbol{i},\qquad &\boldsymbol{ki}=-%
\boldsymbol{ik}=\boldsymbol{j}.
\end{eqnarray}%
Any element of $\mathbb{H}$ can be expanded as $q=q_{0}+q_{1}\boldsymbol{i}%
+q_{2}\boldsymbol{j}+q_{3}\boldsymbol{k}$, where $q_{i}$ ($i=0,1,2,3$) are
real numbers. Quaternion has been used recently as a tool to analyze the 3D
Landau levels \cite{li2013high}.

For our purpose, it is more convenient to write $q$ in the polar-like
coordinate with $q=\rho (\cos \theta +\hat{\mathbf{a}}\sin \theta ),$ where $%
\rho \equiv \left\vert q\right\vert =\sqrt{%
q_{0}^{2}+q_{1}^{2}+q_{2}^{2}+q_{3}^{2}}$ is the norm of $q$, $\theta $ is
the angle, and $\hat{\mathbf{a}}=\hat{a}_{1}\boldsymbol{i}+\hat{a}_{2}%
\boldsymbol{j}+\hat{a}_{3}\boldsymbol{k}$ with $\hat{a}_{1}^{2}+\hat{a}%
_{2}^{2}+\hat{a}_{3}^{2}=1$ is a unity vector denoting the direction in the
imaginary space. From the definition, we immediately get
\begin{equation}
q^{n}=\rho ^{n}(\cos n\theta +\hat{\mathbf{a}}\sin n\theta ).
\label{eq:power of q}
\end{equation}

To construct tight-binding lattice Hamiltonians for the TIs or TSCs, we
typically work in the momentum space. The Hamiltonian coefficients are taken
as components $q_{i}$ of a quaternion $q$, which in general depend on the
momentum through the notation $q_{i}(\mathbf{k})$. In a $d$-dimensional ($%
d=1,2,3$) space, the momentum $\mathbf{k}$ takes values from the Brillouin
zone (BZ) characterized by a $d$-dimensional torus $\mathbb{T}^{d}$. The
norm $\left\vert q\right\vert $ of the quaternion $q$ characterizes the
energy scale (energy gap) of the Hamiltonian, which can be taken as $1$ (the
energy unit) without loss of generality, and the topological space of $q$ is
thus characterized by the sphere $\mathbb{S}^{d}$. The Hamiltonian with $%
q_{i}(\mathbf{k})$ as the coefficients can be considered as a map from the
space $\mathbb{T}^{d}$ to $\mathbb{S}^{d}$. If this map is topologically
nontrivial characterized by a topological invariant (usually taken as the
winding number or Chern number) $\Gamma =1$, geometrically (in the sense of
homotopy) it means that the space $\mathbb{T}^{d}$ wraps around $\mathbb{S}%
^{d}$ one time through the map. Now consider a Hamiltonian where the
coefficients are taken as the components of $q^{n}(\mathbf{k})$. From the
geometric representation of $q^{n}$ in the polar coordinate in Eq.\ (\ref%
{eq:power of q}), if the space $\mathbb{T}^{d}$ wraps around $\mathbb{S}^{d}$
one time through the map $\mathbf{k\rightarrow }q(\mathbf{k})$ with $\Gamma
=1$, it will wrap $\mathbb{S}^{d}$ $n\ $times through the map $\mathbf{%
k\rightarrow }q^{n}(\mathbf{k})$ with $\Gamma =n$. So, by this
quaternion-power mapping, we can construct Hamiltonians for topologically
distinct new phases with arbitrary integer topological invariants. This
serves as our physical intuition to construct tight-binding Hamiltonians for
new topological phases. In the following, we apply this method to
construction of the Hamiltonians for all the 3D topological phases in the
periodic table characterized by the integer group $\mathbb{Z}$, which
include the spin-singlet and the spin-triplet TSCs, the chiral and the Hopf
TIs.

\textit{Spin-singlet TSC (class CI)}--- The topological phases in the
periodic table are classified by three generic symmetries, the time
reversal, the particle-hole exchange (charge conjugation), and the chiral
symmetry, denoted respectively by $T,C,S$ with $S\equiv TC$. These three
symmetries can be used to classify the system even when weak disorder breaks
the lattice translational symmetry. The symmetries $T$ and $C$ are
represented by anti-unitary operators, and $T^{2},C^{2}$ can take values
either $+1$ or $-1$ depending on the effective spin of the system.

Spin singlet TSC is described by a Bogoliubov-de-Gennes (BdG) type of
mean-field Hamiltonian and belongs to the symmetry class CI in the periodic
table, which means the BdG\ Hamiltonian has both $T$ and $C$ symmetries with
$T^{2}=1$ and $C^{2}=-1$. The topological phase is characterized by a
topological invariant $\Gamma _{\text{CI}}$, which takes values from $2%
\mathbb{Z}$ (even integers). Ref.\ \cite{schnyder2009lattice} has proposed a
Hamiltonian in a diamond lattice which realizes a special instance of the
CI\ TSC with $\Gamma _{\text{CI}}=\pm 2$. Here, we construct tight binding
Hamiltonians which can realize all the topologically distinct phases for the
CI\ TSC with arbitrary even integer $\Gamma _{\text{CI}}$ in a simple cubic
lattice. The simplified lattice geometry could be important for an
experimental implementation.

To construct the BdG Hamiltonian in the momentum space, first we define a
quaternion $q$ with the following dependence on the momentum $\mathbf{k}$
\begin{equation}
q=t\cos k_{x}-\boldsymbol{i}(\sin k_{x}+\sin k_{y}+\sin k_{z})+\boldsymbol{j}%
\cos k_{y}+\boldsymbol{k}\cos k_{z},  \label{1}
\end{equation}%
where $t$ is a dimensionless parameter. A family of the BdG Hamiltonians can
be constructed on the 3D cubic lattice with the form $H_{\text{CI}}=\sum_{%
\mathbf{k}}\Psi _{\mathbf{k}}^{\dagger }\mathcal{H}_{\text{CI}}(\mathbf{k}%
)\Psi _{\mathbf{k}}$ in the momentum space, where $\Psi _{\mathbf{k}}=(a_{%
\mathbf{k}\uparrow },b_{\mathbf{k\uparrow }},a_{\mathbf{-k\downarrow }%
}^{\dagger },b_{\mathbf{-k\downarrow }}^{\dagger })^{T}$ denotes the
fermionic mode operators with spin $\uparrow ,\downarrow $ and momentum $%
\mathbf{k}$. The $4\times 4$ Hamiltonian matrix reads
\begin{equation}
\mathcal{H}_{\text{CI}}(\mathbf{k})=\left(
\begin{array}{cc}
\mathbf{m}\cdot \boldsymbol{\sigma } & (q^{n})_{3}\mathbf{I_{2}} \\
(q^{n})_{3}\mathbf{I_{2}} & -\mathbf{m}\cdot \boldsymbol{\sigma }%
\end{array}%
\right) ,  \label{2}
\end{equation}%
where $\mathbf{m}=\left( (q^{n})_{0},(q^{n})_{1},(q^{n})_{2}\right) $ with $%
(q^{n})_{i}$ denoting the $i$th-components of the quaternion $q^{n}$, $%
\mathbf{I_{2}}$ is the $2\times 2$ identity matrix, and $\boldsymbol{\sigma }%
=(\sigma ^{x},\sigma ^{y},\sigma ^{z})$ are the Pauli matrices. Expressed in
the real space, the Hamiltonian $H_{\text{CI}}$ contains spin-singlet $d$%
-wave pairing described by the quaternion component $(q^{n})_{3}$, and has
local hopping and pairing terms up to the $n$th neighboring sites. One can
check that $\mathcal{H}_{\text{CI}}(\mathbf{k})$ indeed has both $T$ and $C$
symmetries (and thus also the chiral symmetry $S=TC$) with $T^{2}=1$ and $%
C^{2}=-1$ (see the supplement for an explicit check \cite{Supplement}).

Now we show that the Hamiltonian $H_{\text{CI}}$ has topologically distinct
phases depending on the parameters $n$ and $t$. For this purpose, we need to
calculate the topological invariant $\Gamma _{\text{CI}}$ for $H_{\text{CI}}$%
. Direct diagonalization of the Hamiltonian $H_{\text{CI}}$ leads to the
energy spectrum $E_{\pm }(\mathbf{k})=\pm \left\vert q^{n}\right\vert =\pm
\rho ^{n}=\pm \lbrack t^{2}\cos ^{2}k_{x}+\cos ^{2}k_{y}+\cos
^{2}k_{z}+(\sin k_{x}+\sin k_{y}+\sin k_{z})^{2}]^{n/2}$. It is always
gapped if $t\neq 0$ and has a twofold degeneracy for each $\mathbf{k}$. To
calculate the topological index $\Gamma _{\text{CI}}$, we first flatten the
bands of $H_{\text{CI}}$ (which is a continuous transformation that does not
change its topological property) by introducing the $Q$ matrix,
\begin{equation}
Q(\mathbf{k})=1-2P(\mathbf{k}),\quad P(\mathbf{k})=\sum_{f}|u_{f}(\mathbf{k}%
)\rangle \langle u_{f}(\mathbf{k})|,  \label{3}
\end{equation}%
where $P(\mathbf{k})$ is the projector onto the filled Bloch bands (with
energy $E_{-}(\mathbf{k})$ and wave-vectors $|u_{f}(\mathbf{k})\rangle $
from the diagonalization of $\mathcal{H}_{\text{CI}}$). With the chiral
symmetry, the $Q$ matrix can be brought into the block off-diagonal form $Q(%
\mathbf{k)}=\left(
\begin{array}{cc}
0 & b(\mathbf{k}) \\
b^{\dagger }(\mathbf{k}) & 0%
\end{array}%
\right) $ by a unitary transformation, with
\begin{equation}
b(\mathbf{k})=-\left(
\begin{array}{cc}
(q^{n})_{3}-i(q^{n})_{2} & -i(q^{n})_{0}-(q^{n})_{1} \\
-i(q^{n})_{0}+(q^{n})_{1} & (q^{n})_{3}+i(q^{n})_{2}%
\end{array}%
\right) /E_{+}(\mathbf{k})  \label{8}
\end{equation}%
for the Hamiltonian $H_{\text{CI}}$. With the matrix $b(\mathbf{k})$, the
topological index $\Gamma _{\text{CI}}$ is defined by the following winding
number \cite{ryu2010topological}:
\begin{equation}
\Gamma _{\text{CI}}=\frac{1}{24\pi ^{2}}\int_{\text{BZ}}d\mathbf{k}%
\;\epsilon ^{\mu \rho \lambda }\text{Tr}[(b^{-1}\partial _{\mu
}b)(b^{-1}\partial _{\rho }b)(b^{-1}\partial _{\lambda }b)],
\label{Chiral-invariant}
\end{equation}%
where $\epsilon ^{\mu \rho \lambda }$ is the antisymmetric Levi-Civita
symbol and $\partial _{\mu }b\equiv \partial _{k_{\mu }}b(\mathbf{k})$. When
$n=1$, the integral in $\Gamma _{\text{CI}}$ can be calculated analytically
and we find $\Gamma _{\text{CI}}\left( n=1\right) =2$sign$(t)=\pm 2$. In
general cases, due to the geometric interpretation of the map $q^{n}$, we
immediately get%
\begin{equation}
\Gamma _{\text{CI}}[\mathcal{H}_{\text{CI}}]=2n\text{sign}(t)=\pm 2n.
\label{CI-invariant}
\end{equation}%
This result is confirmed through direct numerical calculations. We integrate
Eq.\ (\ref{Chiral-invariant}) numerically through discretization of the
Brillouin zone. The calculation results for different $n$ are shown in Fig.
1(a), which quickly converge to the exact results in Eq.\ (\ref{CI-invariant}%
) as the number of integration grids increases. As one varies $n$, it is
evident from Eq.\ (\ref{CI-invariant}) that we can realize all the
spin-singlet TSC phases in the CI class through our constructed Hamiltonian $%
H_{\text{CI}}$ with the topological index $\Gamma _{\text{CI}}$ taking
arbitrary even integers.

\textit{Spin-triplet TSC (class DIII)} --- Spin triplet TSCs are described
by the BdG Hamiltonians that have both $T$ and $C$ symmetries with $T^{2}=-1$
and $C^{2}=1$. It belongs to the symmetry class DIII in the periodic table.
The $^{3}\text{He }$superfluid B phase is a well known example in this class
\cite{kitaev2009periodic,ryu2010topological}, but it is not described by a
simple lattice model. Tight-binding lattice Hamiltonians have been
constructed for the DIII-class spin triplet TSCs with the topological index $%
\Gamma _{\text{DIII}}=\pm 1$ \cite%
{fu2010odd,yan2010theoretical,sato2010topological}. Here, we use the
quaternion method to construct tight-binding Hamiltonians for the DIII-class
TSCs with arbitrary integer topological indices $\Gamma _{\text{DIII}}$ in a
simple cubic lattice.

To construct the Hamiltonian, we define a quaternion $q\left( \mathbf{k}%
\right) $ with the following dependence on $\mathbf{k}$
\begin{equation}
q=h+\cos k_{x}+\cos k_{y}+\cos k_{z}+\boldsymbol{i}t\sin k_{x}+\boldsymbol{j}%
\sin k_{y}+\boldsymbol{k}\sin k_{z}.  \label{qua-DIII}
\end{equation}%
with $t,h$ being dimensionless parameters. We will use this form of $q\left(
\mathbf{k}\right) $ for all our following examples. We construct a four-band
BdG Hamiltonian with the form $H_{\text{DIII}}=\sum_{\mathbf{k}}\Phi _{%
\mathbf{k}}^{\dagger }\mathcal{H}_{\text{DIII}}(\mathbf{k})\Phi _{\mathbf{k}}
$, with the fermionic mode operators $\Phi _{\mathbf{k}}=(a_{\mathbf{%
k\uparrow }},a_{\mathbf{k}\downarrow },a_{-\mathbf{k}\uparrow }^{\dagger
},a_{-\mathbf{k}\downarrow }^{\dagger })^{T}$ and the $4\times 4$
Hamiltonian matrix
\begin{equation}
\mathcal{H}_{\text{DIII}}(\mathbf{k})=\mathbf{u}\cdot \boldsymbol{\Gamma },
\label{Ham-DIII}
\end{equation}%
where $\mathbf{u}=((q^{n})_{1},(q^{n})_{2},(q^{n})_{3},(q^{n})_{0})$, $%
\boldsymbol{\Gamma }=(\gamma ^{0}\gamma ^{1},\gamma ^{0}\gamma ^{2},\gamma
^{0}\gamma ^{3},-i\gamma ^{0}\gamma ^{5})$, and $\gamma ^{i}$ denote the
standard Dirac matrices with the explicit expressions given in the
supplement \cite{Supplement}. This Hamiltonian has spin triplet pairing with
the energy spectrum $E_{\pm }(\mathbf{k})=\pm \left\vert \mathbf{u}(\mathbf{k%
})\right\vert =\pm \left\vert q(\mathbf{k})\right\vert ^{n}$, which is fully
gapped when $|h|\neq 1,3$ and $t\neq 0$.

The DIII class TSC has the chiral symmetry, so its $Q$ matrix for the
Hamiltonian can be brought into the block off-diagonal form \cite{Supplement}
and the topological index $\Gamma _{\text{DIII}}$ is also characterized by
the winding number in Eq.\ \eqref{Chiral-invariant}. We find
\begin{equation}
\Gamma _{\text{DIII}}[\mathcal{H}_{\text{DIII}}]=%
\begin{cases}
-2n\text{sign}(t) & |h|<1 \\
n\text{sign}(t) & 1<|h|<3 \\
0 & |h|>3%
\end{cases}%
.  \label{9}
\end{equation}%
It is evident that the topological index $\Gamma _{\text{DIII}}$ can take
arbitrary integer values for our constructed Hamiltonian depending on the
parameters $n,t,h$. In the particular case with $n=t=1$, the Hamiltonian
reduces to the model Hamiltonian introduced in Ref.\ \cite%
{schnyder2008classification,ryu2010topological}, which has $\Gamma _{\text{%
DIII}}=1$ or $-2$.

\textit{Chiral TI (class AIII)}--- Chiral TIs do not have time-reversal or
particle-hole symmetry (thus $T=C=0$), but they possess chiral symmetry with
$S=1$ and belongs to the symmetry class AIII in the periodic table.
Tight-binding Hamiltonians have been constructed for the chiral TIs with the
topological index $\Gamma _{\text{AIII}}=\pm 1$ \cite%
{neupert2012noncommutative}. Here, we use the quaternion method to construct
Hamiltonians with arbitrary integer $\Gamma _{\text{AIII}}$. We consider a
three-band Hamiltonian with the following form $H_{\text{AIII}}=\sum_{%
\mathbf{k}}\xi _{\mathbf{k}}^{\dagger }\mathcal{H}_{\text{AIII}}(\mathbf{k}%
)\xi _{\mathbf{k}}$, where the fermionic mode operators $\xi _{\mathbf{k}%
}=(a_{\mathbf{k}},b_{\mathbf{k}},c_{\mathbf{k}})^{T}$ and the $3\times 3$
Hamiltonian matrix
\begin{equation}
\mathcal{H}_{\text{AIII}}(\mathbf{k})=\mathbf{u}\cdot \mathcal{G}.
\label{10}
\end{equation}%
In $\mathcal{H}_{\text{AIII}}$, $\mathbf{u}$ denotes the same quaternion
coefficients as defined below Eq.\ \eqref{Ham-DIII} and $\mathcal{G}%
=(\lambda _{4},\lambda _{5},\lambda _{6},\lambda _{7})$ are the four
Gell-Mann matrices with the explicit form given in the supplement \cite%
{Supplement}. The Hamiltonian $H_{\text{AIII}}$ is gapped when $|h|\neq 1,3$
and $t\neq 0$ and has a perfectly flat middle band with a macroscopic number
of zero-energy modes due to the chiral symmetry \cite%
{neupert2012noncommutative}. A topological invariant classifying this family
of Hamiltonians can be defined as \cite{neupert2012noncommutative}
\begin{equation}
\Gamma _{\text{AIII}}=-\frac{1}{12\pi ^{2}}\int_{\text{BZ}}d\mathbf{k}%
\;\epsilon ^{\alpha \beta \gamma \rho }\epsilon ^{\mu \nu \tau }\frac{1}{|%
\mathbf{u}|^{4}}\bm{\mathbf{u}}_{\alpha }\partial _{\mu }\bm{\mathbf{u}}%
_{\beta }\partial _{\nu }\bm{\mathbf{u}}_{\gamma }\partial _{\tau }%
\bm{\mathbf{u}}_{\rho }.  \label{11}
\end{equation}%
We have calculated this invariant and found that
\begin{equation}
\Gamma _{\text{AIII}}[\mathcal{H}_{\text{AIII}}]=n\text{sign}(t)=\pm n,\text{
}\left( 1<|h|<3\right)   \label{12}
\end{equation}%
for our constructed $H_{\text{AIII}}$. This analytic result is confirmed
with direct numerical calculations as shown in Fig. 1(b). In the particular
case with $n=1$, the Hamiltonian $H_{\text{AIII}}$ reduces to the model
Hamiltonian constructed in Ref. \cite{neupert2012noncommutative}. Through
the quaternion power, we extend the model Hamiltonian and realize the chiral
TIs with the topological index taking arbitrary integer values.

\begin{figure}[tbp]
(a) \includegraphics[width=0.48\textwidth]{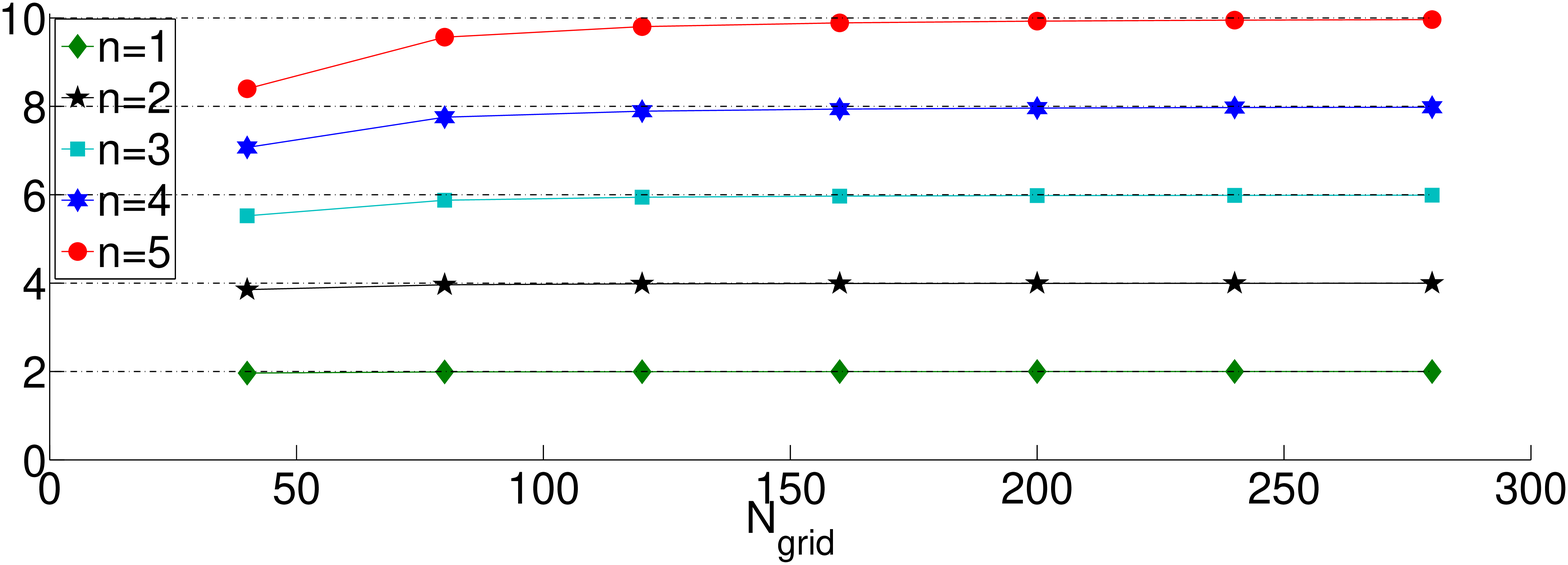} (b) %
\includegraphics[width=0.48\textwidth]{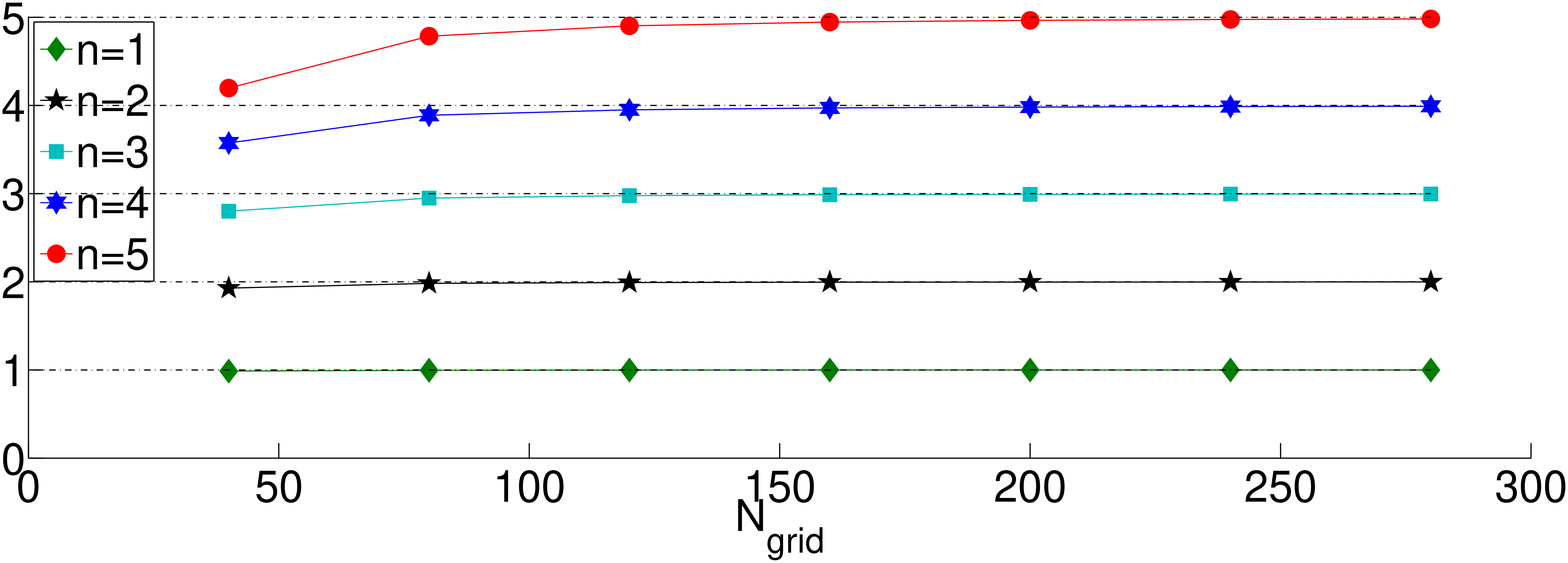}
\caption{Plots of the numerical calculation results for the topological
indices associated with the Hamiltonians in the symmetry class CI (a) and
AIII (b) under different parameter $n$. The topological invariants converge
rapidly to $2n$ in (a) and $n$ in (b) as the number of grids increases in
the discretization. The parameters $t$ and $h$ are chosen as $(t,h)=(1,2)$
\protect\cite{Supplement}.}
\label{fig:Plot-of-the}
\end{figure}

\textit{Hopf insulators (class A)}--- The Hamiltonians in class A do not
have any symmetry ($T$, $C$, or $S$) except the particle number
conservation. Generically, this class of Hamiltonians have no topologically
nontrivial phase in 3D, but there is a peculiar exception, called the Hopf
insulator, which occurs when the Hamiltonian has just two bands due to
existence of the topologically nontrivial Hopf map from $\mathbb{S}^{3}$ to $%
\mathbb{S}^{2}$ \cite{moore2008topological,deng2013hopf}. The Hopf
insulators are characterized by the topological Hopf index $\Gamma _{\text{H}%
}$, which takes values from the integer set $\mathbb{Z}$. A model
Hamiltonian has been constructed for the Hopf insulator with $\Gamma _{\text{%
H}}=\pm 1$ in Ref. \cite{moore2008topological} based on the Hopf map \cite%
{whitehead1947expression}. This method was extended in Ref. \cite%
{deng2013hopf} to construct Hamiltonians for general Hopf insulators with
arbitrary integer $\Gamma _{\text{H}}$ based on the generalized Hopf map
encountered in mathematics but not in physics literature. Here, with the
quaternion algebra, we use only the simple Hopf map but still can construct
tight-binding Hamiltonians for the Hopf insulators with arbitrary integer $%
\Gamma _{\text{H}}$.

To construct the Hamiltonian, we define two complex variables $\eta =(\eta
_{\uparrow },\eta _{\downarrow })^{\text{T}}$ from the quaternion $\eta
_{\uparrow }=(q^{n})_{1}+i(q^{n})_{2},\;\eta _{\downarrow
}=(q^{n})_{3}+i(q^{n})_{0}$, where $q\left( \mathbf{k}\right) $ is defined
by Eq.\ \eqref{qua-DIII}. The Hopf map is defined as $\mathbf{v}=\eta
^{\dagger }\boldsymbol{\sigma }\eta $, which is a quadratic map from $%
\mathbb{S}^{3}\rightarrow \mathbb{S}^{2}$ up to normalization. The two-band
Hamiltonians can then be constructed as $H_{\text{Hopf}}=\sum_{\mathbf{k}%
}\psi _{\mathbf{k}}^{\dagger }\mathcal{H}_{\text{Hopf}}(\mathbf{k})\psi _{%
\mathbf{k}}$ with $\psi _{\mathbf{k}}=(a_{\mathbf{k}\uparrow },a_{\mathbf{k}%
\downarrow })^{T}$ and
\begin{equation}
\mathcal{H}_{\text{Hopf}}(\mathbf{k})=\mathbf{v}\cdot \boldsymbol{\sigma }.
\label{13}
\end{equation}%
The Hopf insulators are characterized by the topological Hopf index, define
as
\begin{equation}
\Gamma _{\text{H}}[\mathcal{H}_{\text{Hopf}}]=-\int_{\text{BZ}}\mathbf{%
F\cdot \mathbf{A}}\;d^{3}\mathbf{k},  \label{14}
\end{equation}%
where $\mathbf{F}$ is the Berry curvature with\ $F_{\mu }\equiv \frac{1}{%
8\pi }\epsilon _{\mu \nu \tau }\mathbf{v}\cdot (\partial _{\nu }\mathbf{%
v\times \partial _{\tau }\mathbf{v}})$ and $\mathbf{A}$ is the associated
Berry connection which satisfies $\nabla \times \mathbf{A}=\mathbf{F}$ \cite%
{moore2008topological,deng2013hopf}. From this definition and our geometric
interpretation of $q^{n}$, we find
\begin{equation}
\Gamma _{\text{H}}[\mathcal{H}_{\text{Hopf}}]=n\text{sign}(t)=\pm n,\quad
(1<|h|<3)
\end{equation}%
This analytical expression is also confirmed with direct numerical
calculations. Some numerical results for the topological indices are listed
in Table (\ref{tab:Numerical-calculations-of}) for different classes of TIs
and TSCs, which agree very well with our analytical expressions.

\begin{table}[tbp]
\includegraphics[width=0.45\textwidth]{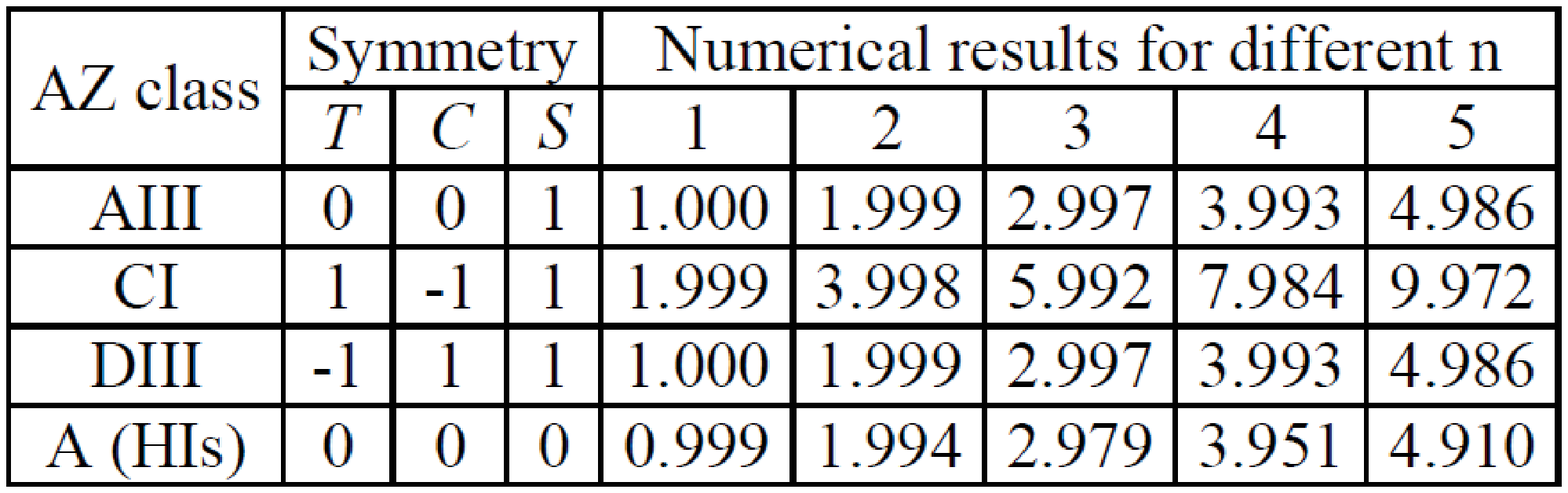}
\caption{Numerical results of the corresponding topological invariants for
the constructed Hamiltonians in different symmetry classes. The symmetry
property of each class is also indicated. The presence of time reversal
symmetry $T $, particle-hole symmetry $C$, and chiral symmetry $S $ is
denoted by $\pm 1,$ with $\pm 1$ specifying the values of $T ^{2}$ and $C
^{2}$. The absence of these symmetries is denoted by $0.$ The parameters for
the corresponding Hamiltonians are chosen as $(t,h)=(1,2)$. The number of
grid points are $N_{\text{grid}}=320$ for all the cases \protect\cite%
{Supplement}. }
\label{tab:Numerical-calculations-of}
\end{table}

Before ending the paper, we briefly remark that the quaternion tool proposed
here can be extended straightforwardly to the 1D and 2D cases although our
focus in this paper is on the 3D topological phases. We can set one (two) of
the quaternion components to zero for the 2D (1D) case and observe that the
map $q\longrightarrow q^{n}$ always preserves the subspace of $\mathbb{H}$
spanned by $\{1,\boldsymbol{i},\boldsymbol{j}\}$) ($\{1,\boldsymbol{i}\}$).
With the power mapping $q^{n}$ which preserve the symmetry of the
Hamiltonian, starting from one particular example of topological
Hamiltonians with the topological index $\Gamma =\pm 1$, we can always
construct a family of Hamiltonians which realize all the topological phases
with arbitrary integer $\Gamma $.

In summary, we have proposed a powerful tool based on the quaternion algebra
to systematically construct tight-binding Hamiltonians for all the
topological phases in the periodic table that are characterized by arbitrary
integer topological indices. The constructed Hamiltonians make the basis for
further studies of properties of these topological phases and phase
transitions and provide an important step for future experimental
realization.

\begin{acknowledgments}
We thank K. Sun and C. J. Wu for helpful discussions. This work was
supported by the NBRPC (973 Program) 2011CBA00300 (2011CBA00302), the IARPA
MUSIQC program, the ARO and the AFOSR MURI program.
\end{acknowledgments}

%


\onecolumngrid
\appendix
\clearpage
\section*{Supplementary Material: Systematic Construction of tight-binding
Hamiltonians for Topological Insulators and Superconductors}

\begin{quote}
In this supplementary material, we explicitly check the symmetries for our
constructed Hamiltonians. We also give some details for the description of these
Hamiltonians and the calculation of their corresponding topological indices.
\end{quote}

We first specify the definition of time-reversal ($T$), particle-hole
(charge conjugation $C$), and chiral ($S=TC$) symmetries in the momentum ($%
\mathbf{k}$) space. A Hamiltonian is represented by a finite matrix $%
\mathcal{H}(\mathbf{k})$ in the $\mathbf{k}$ space (kernel of the Hamiltonian). It has the time reversal
symmetry if there exists a unitary matrix $T_{m}$ such that
\begin{equation}
T_{m}\mathcal{H}^{\ast }(\mathbf{k})T_{m}^{-1}=\mathcal{H}(-\mathbf{k}).
\label{1}
\end{equation}%
Similarly, $\mathcal{H}(\mathbf{k})$ has the particle-hole symmetry if
there is a unitary matrix $C_{m}$ such that
\begin{equation}
C_{m}\mathcal{H}^{\ast }(\mathbf{k})C_{m}^{-1}=-\mathcal{H}(-\mathbf{k}).
\label{2}
\end{equation}%
The anti-unitary nature of the time reversal and the particle-hole symmetries is manifested in the complex conjugate $\mathcal{H}^{\ast }(\mathbf{k})$ in Eqs.\ (1) and (2). Finally, as $S=TC$, the chiral symmetry is unitary and
represented by ($S_{m}=T_{m}C_{m}^{\ast }$)
\begin{equation}
S_{m}\mathcal{H}(\mathbf{k})S_{m}^{-1}=-\mathcal{H}(\mathbf{k}).  \label{3}
\end{equation}
Pertaining to the presence/absence of these symmetries, ten classes of single-particle Hamiltonians can be specified, which is intimately related to the classification of random matrices by Altland and Zirnbauer (AZ) \cite{altland1997nonstandard}. 

\section{Spin-singlet topological superconductor}

We first prove that the Hamiltonian $H_{\text{CI}}$ for the spin-singlet
topological superconductors (TSCs) constructed in Eq.\ (4) of the main text
has the CI-class symmetry with $T^{2}=1$ and $C^{2}=-1$. Let us denote the
parity of a function $f(\mathbf{k})$ by $P[f(\mathbf{k})]$, with $P=1$ ($-1$%
) for an even (odd) parity under the exchange $\mathbf{k}\rightarrow -%
\mathbf{k}$. From the product table of the quaternion algebra, it is easy to
check that with $P[(q)_{1}]=P[(\sin k_{x}+\sin k_{y}+\sin k_{z})]=-1$ and $%
P[(q)_{0}]=P[(q)_{2}]=P[(q)_{3}]=1$, we have $%
P[(q^{n})_{0}]=P[(q^{n})_{2}]=P[(q^{n})_{3}]=1$ and $P[(q^{n})_{1}]=-1$ for
any integer power $n$. The explicit expression of the Hamiltonian in eq.\ (4) of the main text is
\begin{equation}
\mathcal{H}_{\text{CI}}(\mathbf{k})=\left(
\begin{array}{cccc}
(q^{n})_{2} & (q^{n})_{0}-i(q^{n})_{1} & (q^{n})_{3} & 0 \\
(q^{n})_{0}+i(q^{n})_{1} & -(q^{n})_{2} & 0 & (q^{n})_{3} \\
(q^{n})_{3} & 0 & -(q^{n})_{2} & -(q^{n})_{0}+i(q^{n})_{1} \\
0 & (q^{n})_{3} & -(q^{n})_{0}-i(q^{n})_{1} & (q^{n})_{2}%
\end{array}%
\right) .  \label{4}
\end{equation}%
From the parity of $q^{n}$, the time reversal symmetry can be readily seen as
\begin{equation}
\lbrack \mathcal{H}_{\text{CI}}(\mathbf{k})]^{\ast }=\mathcal{H}_{\text{CI}%
}(-\mathbf{k}),  \label{5}
\end{equation}%
so $T_{m}=\mathbf{I}_{4}$, the $4\times 4$ identity matrix. The particle-hole
symmetry can be seen as
\begin{equation}
C_{m}[\mathcal{H}_{\text{CI}}(\mathbf{k})]^{\ast }C_{m}^{-1}=-\mathcal{H}_{%
\text{CI}}(-\mathbf{k}),  \label{6}
\end{equation}%
with $C_{m}=\mathbf{I}_{2}\otimes \sigma ^{y}$,
where $\boldsymbol{\sigma }=(\sigma ^{x},\sigma ^{y},\sigma ^{z})$ denote
the Pauli matrices. Apparently, $T^{2}=1$ and $C^{2}=-1$ (as $%
C_{m}C_{m}^{\ast }=-\mathbf{I}_{4}$), as it is the case for the CI-class symmetry.

To calculate the topological invariant, we note that the system also has the
chiral symmetry $S=TC$ and the $Q(\mathbf{k})$ matrix (defined in the main text) can thus
be brought into the block off-diagonal form by a unitary transformation \cite%
{schnyder2008classification,ryu2010topological}. At the half filling
(therefore inside the energy gap) and with a convenient gauge, direct
calculation leads to%
\begin{equation}
Q(\mathbf{k})=\left(
\begin{array}{cc}
0 & b(\mathbf{k}) \\
b^{\dagger }(\mathbf{k}) & 0%
\end{array}%
\right),   \\
\qquad
b(\mathbf{k})=-\left(
\begin{array}{cc}
(q^{n})_{3}-i(q^{n})_{2} & -i(q^{n})_{0}-(q^{n})_{1} \\
-i(q^{n})_{0}+(q^{n})_{1} & (q^{n})_{3}+i(q^{n})_{2}%
\end{array}%
\right) /E_{+}(\mathbf{k})  \label{8}
\end{equation}
with $E_{+}(\mathbf{k})=\left\vert q(\mathbf{k})\right\vert ^{n}$, as mentioned in the main text.

\section{Spin-triplet topological superconductor}

The Dirac matrices (also known as the gamma matrices) $\{\gamma ^{0},\gamma
^{1},\gamma ^{2},\gamma ^{3}\}$ are a set of $4\times 4$ matrices, defined
as
\begin{equation*}
\gamma ^{0}=\left(
\begin{array}{cc}
\mathbf{I}_{2} & \mathbf{0} \\
\mathbf{0} & -\mathbf{I}_{2}%
\end{array}%
\right) ,\quad \gamma ^{1}=\left(
\begin{array}{cc}
\mathbf{0} & \sigma ^{x} \\
-\sigma ^{x} & \mathbf{0}%
\end{array}%
\right) ,\quad \gamma ^{2}=\left(
\begin{array}{cc}
\mathbf{0} & \sigma ^{y} \\
-\sigma ^{y} & \mathbf{0}%
\end{array}%
\right) ,\quad \gamma ^{3}=\left(
\begin{array}{cc}
\mathbf{0} & \sigma ^{z} \\
-\sigma ^{z} & \mathbf{0}%
\end{array}%
\right) ,
\end{equation*}%
The fifth gamma matrix is defined by $\gamma ^{5}=i\gamma ^{0}\gamma
^{1}\gamma ^{2}\gamma ^{3}=\sigma ^{x}\otimes \mathbf{I}_{2}.$ Using the
explicit form of these gamma matrices, the Hamiltonian matrix $\mathcal{H}_{%
\text{DIII}}(\mathbf{k})$ can be written as
\begin{equation}
\mathcal{H}_{\text{DIII}}(\mathbf{k})=
\left(\begin{array}{cccc}0 & 0 & -i(q^{n})_{0}+(q^{n})_{3}  &  (q^{n})_{1}-i(q^{n})_{2} \\0 & 0 & (q^{n})_{1}+i(q^{n})_{2} & -i(q^{n})_{0}-(q^{n})_{3} \\i(q^{n})_{0}+(q^{n})_{3}  &  (q^{n})_{1}+i(q^{n})_{2} & 0 & 0 \\ (q^{n})_{1}-i(q^{n})_{2} & i(q^{n})_{0}-(q^{n})_{3}  & 0 & 0\end{array}\right).
\end{equation}%
The $Q(\mathbf{k})$ matrix can thus be written as
\begin{equation}
Q(\mathbf{k})=\left(
\begin{array}{cc}
0 & b(\mathbf{k}) \\
b^{\dagger }(\mathbf{k}) & 0%
\end{array}%
\right),   \\
\qquad
b(\mathbf{k})=-\left(
\begin{array}{cc}
-i(q^{n})_{0}+(q^{n})_{3} & (q^{n})_{1}-i(q^{n})_{2} \\
(q^{n})_{1}+i(q^{n})_{2} & -i(q^{n})_{0}-(q^{n})_{3}%
\end{array}%
\right) /E_{+}(\mathbf{k}).  \label{10}
\end{equation}
with $E_{+}(\mathbf{k})=\left\vert q(\mathbf{k})\right\vert ^{n}$. Note that for the quaternion $q$ defined in Eq.\ (9) of the main text, we
have $P[(q)_{1}]=P[(q)_{2}]=P[(q)_{3}]=-P[(q)_{0}]=-1$ and thus  $%
P[(q^{n})_{1}]=P[(q^{n})_{2}]=P[(q^{n})_{3}]=-P[(q^{n})_{0}]=-1$. With the parity properties, one can easily check that
\begin{eqnarray}
(\sigma _{y}\otimes \sigma _{x})[\mathcal{H}_{\text{DIII}}(\mathbf{k}%
)]^{\ast }(\sigma _{y}\otimes \sigma _{x}) &=&\mathcal{H}_{\text{DIII}}(-%
\mathbf{k}),  \label{11} \\
(\sigma _{y}\otimes \sigma _{y})[\mathcal{H}_{\text{DIII}}(\mathbf{k}%
)]^{\ast }(\sigma _{y}\otimes \sigma _{y}) &=&-\mathcal{H}_{\text{DIII}}(-%
\mathbf{k}).  \label{12}
\end{eqnarray}%
So the symmetry matrix $T_{m}=\sigma _{y}\otimes \sigma _{x}$ and $%
C_{m}=\sigma _{y}\otimes \sigma _{y}$ with $T^{2}=-1$ and $C^{2}=1$ (as $%
T_{m}T_{m}^{\ast }=-\mathbf{I}_{4}$ and $C_{m}C_{m}^{\ast }=\mathbf{I}_{4}$%
), as it is the case for the DIII-class symmetry.

\section{Chiral topological insulator}

The four Gell-Mann matrices used in the text are defined as
\begin{equation}
\lambda _{4}=\left(
\begin{array}{ccc}
0 & 0 & 1 \\
0 & 0 & 0 \\
1 & 0 & 0%
\end{array}%
\right) ,\quad \lambda _{5}=\left(
\begin{array}{ccc}
0 & 0 & -i \\
0 & 0 & 0 \\
i & 0 & 0%
\end{array}%
\right) ,\quad \lambda _{6}=\left(
\begin{array}{ccc}
0 & 0 & 0 \\
0 & 0 & 1 \\
0 & 1 & 0%
\end{array}%
\right) ,\quad \lambda _{7}=\left(
\begin{array}{ccc}
0 & 0 & 0 \\
0 & 0 & -i \\
0 & i & 0%
\end{array}%
\right) .  \label{13}
\end{equation}%
The Hamiltonian matrix $\mathcal{H}_{\text{AIII}}(\mathbf{k})$ for the
chiral topological insulator has the following explicit form
\begin{equation}
\mathcal{H}_{\text{AIII}}(\mathbf{k})=\left(
\begin{array}{ccc}
0 & 0 & (q^{n})_{1}-i(q^{n})_{2} \\
0 & 0 & (q^{n})_{3}-i(q^{n})_{0} \\
(q^{n})_{1}+i(q^{n})_{2} & (q^{n})_{3}+i(q^{n})_{0} & 0%
\end{array}%
\right) .  \label{14}
\end{equation}%
The Hamiltonian $\mathcal{H}_{\text{AIII}}(\mathbf{k})$ does not have
time-reversal or particle-hole symmetry, but it has a chiral symmetry $S_{m}%
\mathcal{H}_{\text{AIII}}(\mathbf{k})S_{m}^{-1}=-\mathcal{H}_{\text{AIII}}(%
\mathbf{k})$ with the unitary matrix
\begin{equation}
S_{m}=\left(
\begin{array}{ccc}
1 & 0 & 0 \\
0 & 1 & 0 \\
0 & 0 & -1%
\end{array}%
\right) .  \label{15}
\end{equation}

\end{document}